\documentclass[aps,prl,twocolumn,showpacs,psfig,superscriptaddress]{revtex4-1}
\usepackage{times}
\usepackage{graphicx}
\usepackage{float}
\usepackage{latexsym,amsmath,amssymb,bm,euscript}
\usepackage{color}
\usepackage{subfigure}
\usepackage{epstopdf}
\usepackage[colorlinks=true,linkcolor=blue,citecolor=blue]{hyperref}
\usepackage{hyperref}
\usepackage{soul}
\usepackage{ulem}
\usepackage{mathrsfs}
\usepackage{amsmath}

\begin{document}
\title{Conformal Thermal Tensor Network and Universal Entropy on Topological Manifolds}

\author{Lei Chen}
\affiliation{Department of Physics, Key Laboratory of Micro-Nano Measurement-Manipulation and Physics (Ministry of Education), Beihang University, Beijing 100191, China}

\author{Hao-Xin Wang}
\affiliation{Department of Physics, Key Laboratory of Micro-Nano Measurement-Manipulation and Physics (Ministry of Education), Beihang University, Beijing 100191, China}

\author{Lei Wang}
\affiliation{Institute of Physics, Chinese Academy of Sciences, P.O. Box 603, Beijing 100190, China}


\author{Wei Li}
\email{w.li@buaa.edu.cn}
\affiliation{Department of Physics, Key Laboratory of Micro-Nano Measurement-Manipulation and Physics (Ministry of Education), Beihang University, Beijing 100191, China}
\affiliation{International Research Institute of Multidisciplinary Science, Beihang University, Beijing 100191, China}

\begin{abstract}
Partition functions of quantum critical systems, expressed as conformal thermal tensor networks,  are defined on various manifolds which can give rise to universal entropy corrections. Through high-precision tensor network simulations of several quantum chains, we identify the universal entropy $S_{\mathcal{K}} = \ln{k}$ on the Klein bottle, where $k$ relates to quantum dimensions of the primary fields in conformal field theory (CFT). Different from the celebrated Affleck-Ludwig boundary entropy $\ln{g}$ ($g$ reflects non-integer groundstate degeneracy), $S_{\mathcal{K}}$ has \textit{no} boundary dependence or surface energy terms accompanied, and can be very conveniently extracted from thermal data. On the M\"obius-strip manifold, we uncover an entropy $S_{\mathcal{M}} = \frac{1}{2} (\ln{g} + \ln{k})$ in CFT, where $\frac{1}{2} \ln{g}$ is associated with the only open edge of the M\"obius strip, and $\frac{1}{2} \ln{k}$ with the non-orientable topology. We employ $S_{\mathcal{K}}$ to accurately pinpoint the quantum phase transitions, even for those without local order parameters.
\end{abstract}
\maketitle

{\textit{Introduction}}.--- Quantum critical point (QCP) expands to a finite regime at $T>0$, where the thermal properties exhibit intriguing universal features \cite{Sachdev-QPT}. For $d$ dimensional quantum critical systems, there emerges conformal symmetry in the partition function defined on $d+1$ manifold (Euclidean worldsheet), described by the conformal field theory (CFT)  \cite{CFT, Cardy-CFT}. The partition functions can be expressed as thermal tensor networks (TTNs) \cite{LTRG,LTRG++, SETTN}, which provides a powerful tool exploring finite-temperature properties of exotic thermal matter near QCPs, e.g., extracting universal data to identify the corresponding CFTs.

\begin{figure}[!tbp]
  \includegraphics[angle=0,width=1\linewidth]{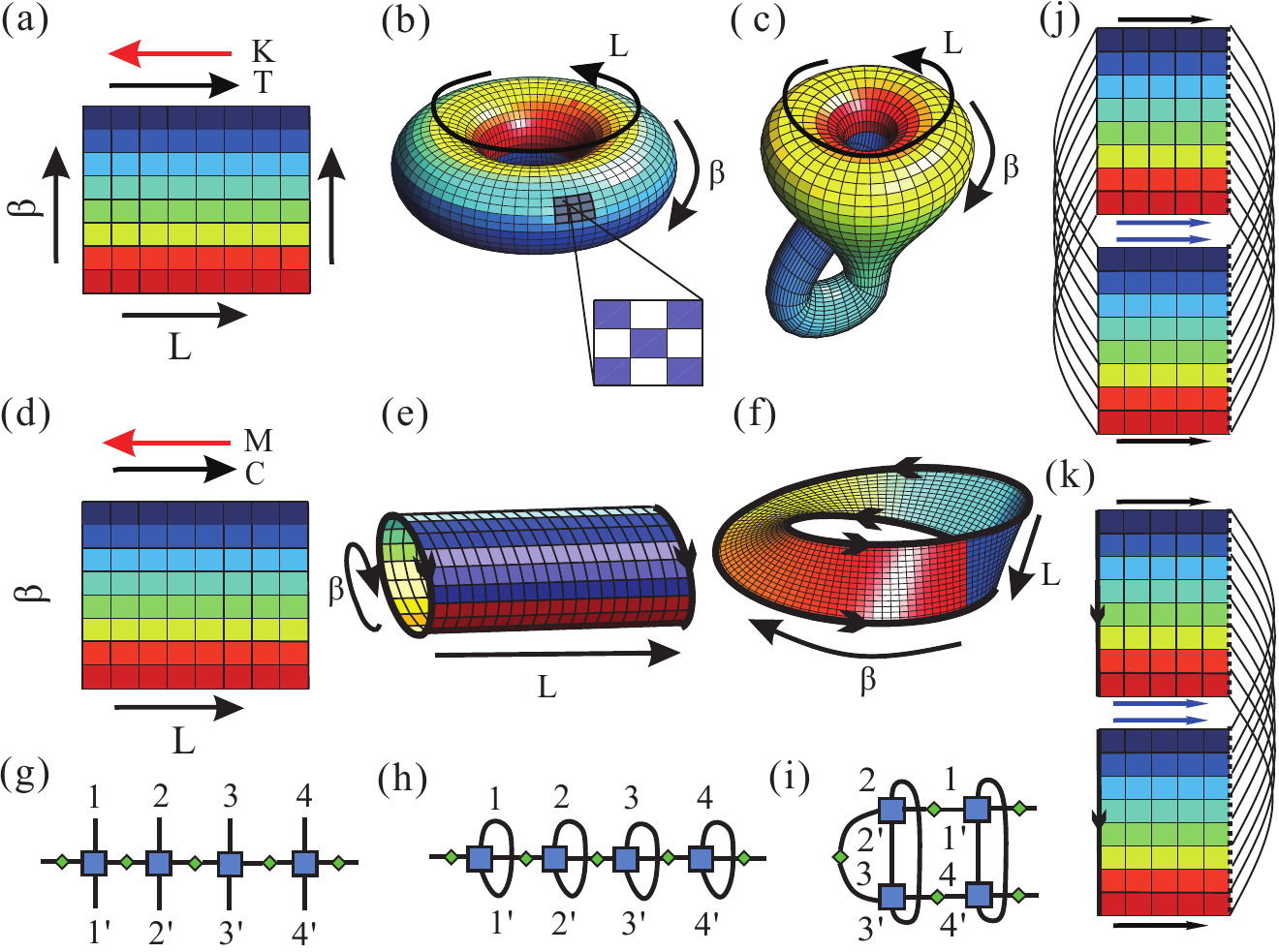}
  \caption{(Color online) (a) TTNs on various 2D manifolds with edges glued together to match the arrows. In spatially PBC cases, the parallel and antiparallel horizontal arrows lead to (b) the torus and (c) the Klein botte, respectively; (d) has OBC in space, where the parallel and antiparallel horizontal arrows correspond to (e) the cylinder and (f) the M\"obius-strip topologies, respectively. (g) represents an MPO, whose (h) direct trace produces the torus or cylinder results, while the twisted trace (i) results in the Klein or M\"obius partition function. (j) The Klein bottle $(L, \beta)$ is ``cut" along a vertical line and rearranged into a $(\frac{L}{2}, 2\beta)$ TTN (parallel arrows of the same color match), with two crosscap edges. (k) The M\"obius strip is transformed into a flat strip with one open and one crosscap edges.}
  \label{Fig:TNGeom}
\end{figure}

According to 2D CFT, for the quantum critical chain with a torus worldsheet, i.e., periodic boundary condition (PBC) on both $L$ and $\beta$ directions [Figs.~\ref{Fig:TNGeom}(a,b)], the (logarithmic) partition function takes the universal form (when $L\gg v\beta$)
$\ln{\mathcal{Z^T}}=L[-\epsilon_0 \beta + \frac{\pi c}{6 v \beta} + \mathcal{O}(\frac{1}{\beta^2})]$, where $\epsilon_0$ is a non-universal constant and $v$ is the speed of ``light" \cite{Affleck-1986, Blote-1986}. Besides the torus, in the seminal paper Ref. \onlinecite{AL-1991}, Affleck and Ludwig proposed a universal boundary entropy by considering a cylindrical worldsheet with open boundary condition (OBC) in the spatial and PBC in the Euclidean-time axis [Figs. \ref{Fig:TNGeom}(d,e)]. The cylinder partition function is $\ln{\mathcal{Z}^C} \simeq L(-\beta \epsilon_0+\frac{\pi c}{6 v \beta}) + \ln{z_b^c}$, where $\ln{z_b^c} = -2 \beta e_s + S_{\rm{AL}}$ represents a boundary partition function. $e_s$ is a nonuniversal surface energy per boundary, and $S_{\rm{AL}}=\ln(g)$ is the well-established Affleck-Ludwig (AL) entropy \cite{FN:g1g2}. Finding more universal data for characterizing CFTs is always interesting and constitute an important topic in studying QCPs.

Two-dimensional CFT on non-orientable surfaces (like the M\"obius strip, crosscap, and the Klein bottle) have been considered in string theory, where the quotient of the original CFT by the worldsheet parity symmetry, i.e., orientifold, has been discussed \cite{CFT-string}.  In condensed matter theory, the exploration of universal thermal data is mainly restricted on the torus and the cylinder manifolds \cite{Affleck-1986, AL-1991, Friedan-2004}, while non-orientable topologies are rarely discussed \cite{Lu-2001}. Until lately, Tu proposes the existence of a universal entropy in 2D CFT on a Klein bottle \cite{Tu-2017}. 

In this work, we consider critical quantum chains defined on two non-orientable manifolds. We show numerically that there exists on the Klein bottle [see Fig. \ref{Fig:TNGeom}(c)] a universal zero-point entropy $S_{\mathcal{K}} = \ln{k}$, where $k = \sum_{a} d_{a}/\mathcal{D}$ sums over all quantum dimensions $d_{a}$ of CFT primary fields (with $\mathcal{D}= \sqrt{\sum_{a} d_{a}^2}$ the total quantum dimension) \cite{Tu-2017}. We also study the antiferromagnetic (AF) Ising chain where the reflection parity of CFT fields has a significant influence, leading to various Klein entropies related to quantum dimensions of CFT primaries. Intriguingly, on a M\"obius-strip worldsheet [Fig. \ref{Fig:TNGeom}(f)], we disclose a distinct entropy correction $S_{\mathcal{M}} = \frac{1}{2} (S_{\rm{AL}}+S_{\mathcal{K}})$, where the factor ``$\frac{1}{2}$" can be understood from the peculiar topology of a M\"obius strip. Furthermore, we show that the universal entropy can be employed to pinpoint QCPs including the conventional quantum phase transitions, as well those beyond the symmetry-breaking paradigm.

{\textit{Models and tensor network methods}}.--- Three critical spin chains, i.e., the spin-1/2 transverse-field Ising (TFI) model
$H_{\rm{TFI}} = \sum_{i} (J S_i^x S_{i+1}^x - h S_i^z)$, critical at $h_c=0.5$ in either ferromagnetic (FM, $J=-1$) or antiferromagnetic (AF, $J=1$) case; the XY model $H_{\rm{XY}} = - \sum_{i} (S_i^x S_{i+1}^x + S_i^y S_{i+1}^y)$; and the spin-1 Blume-Capel (BC) chain \cite{Blume-1966,Capel-1966} $H_{\rm{BC}} = - \sum_{i} [S_i^z S_{i+1}^z - D^z (S^z_i)^2 - h S_i^x]$, tricritical at $D^z=0.9103, h=0.4155$ \cite{Alcaraz-1985,Balbao}. The quantum criticality of  the TFI model is described by the $c=1/2$ Ising CFT with three primary fields {$\mathbb{I}$, $\sigma$, $\psi$} whose quantum dimensions are $d_{\mathbb{I}}=1, d_{\sigma}=\sqrt{2}$, and $d_{\psi}=1$, respectively. The XY chain is described by a U(1)$_4$ compactified boson field \cite{Suzuki-1995,Maansson-2013}, four Abelian primary fields are all of dimension one. The spin-1 BC model has an emergent superconformal CFT at the tricritical point, and is with $d_{\mathbb{I}}=s_2, d_{\epsilon}= d_{\epsilon'}=s_1, d_{\epsilon''}=s_2, d_{\sigma}=\sqrt{2}s_1, d_{\sigma'}=\sqrt{2}s_2$ (where $s_1=\frac{\sin{(2\pi/5)}}{\sqrt{5}}, s_2=\frac{\sin{(4\pi/5)}}{\sqrt{5}}$) \cite{CFT}. 

In the following, we employ TTNs on various manifolds to investigate thermal properties near these QCPs. Notably, tensor networks are currently utilized as a versatile and powerful means to extract universal CFT data like the conformal central charge $c$, scaling dimensions $(h,\bar{h})$, and operator-product expansion coefficients \cite{Gu-2009,Evenbly-2016,Hauru-2016,Czech-2016,Yang-2017,Milsted-2017}.  In our simulations, two specific approaches are adopted: the linearized tensor renormalization group (LTRG) \cite{LTRG,LTRG++} and the series-expansion TTN (SETTN) algorithms \cite{SETTN}. 

LTRG, based on a checkerboard world-line TTN [Figs. \ref{Fig:TNGeom}(a,b)], was proposed in Ref. \onlinecite{LTRG} firstly for infinite-size systems and adapted to finite-size chains later \cite{LTRG++}. We cool down the system by successively projecting $e^{-\tau h_{ij}}$ to a matrix product operator (MPO) representation of the density matrix $\rho$ [Fig. \ref{Fig:TNGeom}(g)], where $h_{ij}$ is the local Hamiltonian and $\tau$ is a small discritization slice ($\tau= 0.025$ in the present study). SETTN is a discritization-error-free method exploiting the Taylor expansion $\rho(\beta)=e^{-\beta H} = \sum_{n=0}^{N_{\rm{cut}}} \frac{(-\beta)^n}{n!} H^n$, which can be used to calculate both OBC and PBC chains very conveniently,  by exploiting efficient MPO representations of $H^n$ \cite{SETTN}. In practical calculations, both methods agree very well with each other in all cases explored. Note that although the computation of critical systems is in general challenging, as will be shown below, the thermal states employed in extracting universal data are adequate at relatively high temperatures and thus $\chi \simeq 200 \sim 300$ bond states well suffice. Besides numerics, Ising and XY chains can be solved analytically on the torus and the Klein bottle \cite{Tu-2017}, in addition to which the XY chain on the cylinder and the M\"obius-strip also have closed-form expressions, providing valuable benchmarks and useful complements (see Secs. I and II in Supplementary Material for more details of related TTN algorithms and analytical solutions).

\begin{figure}[tbp]
\setlength{\abovecaptionskip}{5pt}
\includegraphics[angle=0,width=1.0\linewidth]{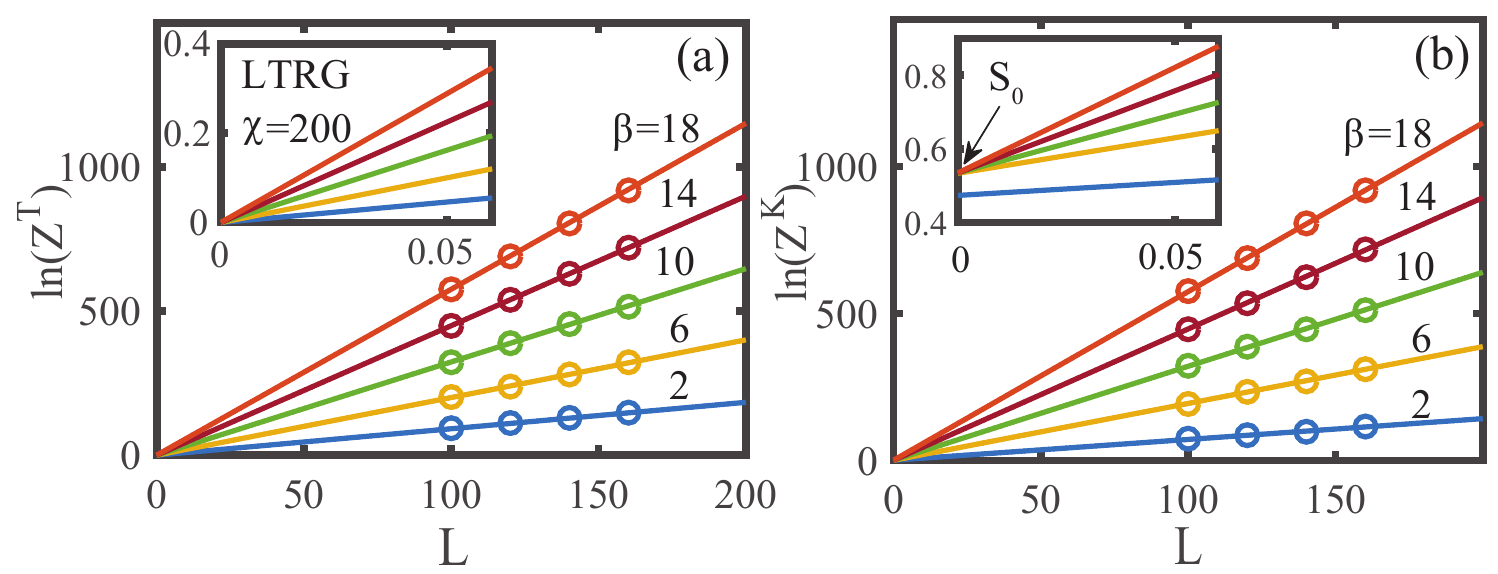}
\caption{(Color online) Linear extrapolations of (a) $\ln{\mathcal{Z}^{\mathcal{T}}}$ on the torus and (b) $\ln{\mathcal{Z}^{\mathcal{K}}}$ on the Klein bottle, based on $L=100,120,140,160$ data. Insets in (a,b) zoom in the regime near $L=0$.} 
\label{Fig:Ising}
\end{figure}

\textit{Spatial reflection and universal Klein entropy}.--- A direct trace $\rm{Tr[\rho(\beta))]}$ results in the torus or the cylinder partition function, which can be implemented by contracting the two physical indices attached to the same site, like 1 with $1^\prime$, 2 with $2^\prime$, etc, as shown in Fig. \ref{Fig:TNGeom}(h). On the contrary, a twisted trace $\rm{Tr[\mathcal{P} \rho(\beta)]}$ corresponds to the Klein-bottle or the M\"obius-strip partition function, where $\mathcal{P}$ is a spatial reflection operation mapping $i \to L-i+1$ in a chain of length $L$. In practice, we fold the MPO from the middle and contract corresponding indices, like 2 with $3^\prime$, and $2^\prime$ with 3, etc, all the way from the center to both ends, as shown in Fig. \ref{Fig:TNGeom}(i).

\begin{figure}[tbp]
\includegraphics[angle=0,width=1\linewidth]{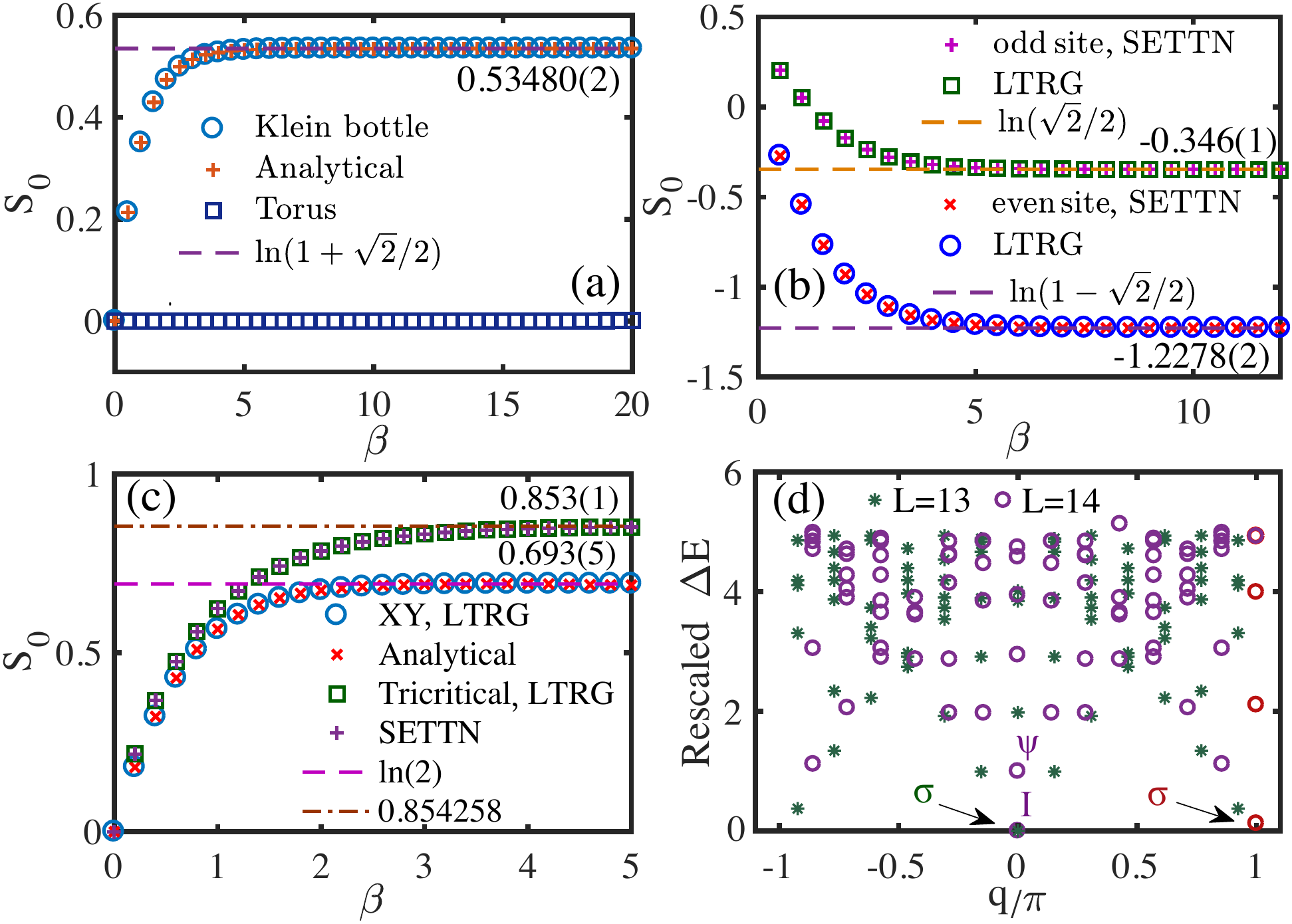}
\caption{(Color online) (a) $S_0$ versus $\beta$, where $S_0=0$ on the torus while $S_0=\ln(1+\frac{\sqrt{2}}{2})$ on the Klein bottle.  (b) $S_0$ of the AF TFI chains with even ($L=50,60,70,80$) and odd ($L=23,35,47,59$) lengths. (c) Universal entropies of the XY and the spin-1 BC chains, simulated on $L=50,60,70,80$ and $L=70,80,90,100$ systems, respectively. (d) The conformal towers are shown for the AF Ising where the odd-parity states at $q=0,\pi$ are marked by red color. } 
\label{Fig:KdiffM}
\end{figure}

We start with the critical TFI chain, and show in Fig. \ref{Fig:Ising}(a) $\ln{\mathcal{Z}^{\mathcal{T}}}$ on the torus, which is linear in $L$ (with gradients varying with $\beta$), and the intercepts are all zero [inset of Fig. \ref{Fig:Ising}(a)]. In Fig. \ref{Fig:Ising}(b), on the contrary, $\ln{\mathcal{Z}^{\mathcal{K}}}$ on the Klein bottle shows a nonzero intercept $S_0$  [inset of Fig. \ref{Fig:Ising}(b)]. $S_0$'s are collected and plotted versus $\beta$ in Fig. \ref{Fig:KdiffM}(a), which very well converge to $0.53480(2) \simeq \ln{(1+\sqrt{2}/2)}$ even at high temperatures ($T/|J|\sim0.2$)! This universal entropy is related to the emergent Ising CFT, where the normalized sum of quantum dimensions is $k=(d_{\mathbb{I}} + d_{\psi} +d_{\sigma})/\mathcal{D}=(2+\sqrt{2})/2$, and the Klein entropy $S_{\mathcal{K}}=\ln{k}$. This motivating observation of nonzero residual entropy can be understood from the Klein-bottle partition function \cite{Tu-2017},

\begin{equation}
\ln{\mathcal{Z^K}} \simeq L (\beta \epsilon_0 + \frac{\pi c}{24v\beta}) + \ln{k}.
\label{Eq:ParFuncKlein}
\end{equation}
Different from the well-known AL entropy which depends on the conformal boundary conditions (and is always accompanied with a surface energy), $S_{\mathcal{K}}$ is easy to be extracted numerically by extrapolations or by estimating ratios in quantum Monte Carlo \cite{Tang-2017}.

\textit{Reflection parity in the AF Ising, XY, and the tricritical Ising models.}--- From the CFT perspective, only nonchiral CFT fields those are symmetric under the spatial relfection $\mathcal{P}$ contribute to (the universal part of) $\mathcal{Z^K}$. The reason has a \textit{topology} origin: if we evolve a left mover along the Euclidean-time direction on the nonorientable worldsheet, it changes to an orthorgonal right mover and the Boltzmann weights does \textit{not} enter the partition function. Therefore, only symmetric states $ |\eta \rangle$ at momentum $q=0$ and $\pi$ contribute (note that $\pi$ and $-\pi$ represent essentially the same lattice momentum). 

In the FM TFI, all relevant fields are at momentum $q=0$ and have the same reflection parity, leading to $S_{\mathcal{K}}=\ln{(1+\frac{\sqrt{2}}{2})}$. However for the AF TFI, in Fig. \ref{Fig:KdiffM}(b), TTN simulations show that $S_0$ converges to $\ln{(1-\sqrt{2}/2)}$ and $\ln{\sqrt{2}/2}$, on chains of even and odd lengths, respectively. To understand these special values of residual entropies, we perform exact diagnalization (ED) and show the rescaled gap $\Delta E$ in Figs. \ref{Fig:KdiffM} (d), where $\Delta E(n,\bar{n}) = \frac{L}{2 \pi v} [E(q)-E_0] = h+\bar{h}+n+\bar{n}$ for $q=2\pi (n-\bar{n})/L$, with $n,\bar{n}$ integers, $h=\bar{h}$ the (holomorphic and antiholomorphic) scaling dimensions \cite{Cardy-1984,Balbao}, and $E_0$ the ground-state energy. For even-length AF chains, we observe that the primary field $\sigma$ and its descendents at $q=\pi$ have odd instead of even parity, which amount to a minus sign before $d_{\sigma}$ in the sum $k^{\prime} = (d_{\mathbb{I}} + d_{\psi} - d_{\sigma})/\mathcal{D}=1-\sqrt{2}/2$ \cite{FN-AFIsing}. As for odd-length, $\mathbb{I}$ and $\psi$ fields move towards (but not equal to) $\pi$ [see Fig. \ref{Fig:KdiffM} (d)], this time only the $\sigma$ field and its descendents at $q=0$ contribute to the twisted partition function, resulting in a $S_0=\ln(d_{\sigma}/\mathcal{D})=\ln{\frac{\sqrt{2}}{2}}$. These two different universal entropies are also remarkable, and from the latter (odd-length AF chain) one directly reads out the quantum dimension of $\sigma$ field, as $d_{\sigma}=\sqrt{2}$.

In Fig. \ref{Fig:KdiffM}(b), we present the results of $S_{\mathcal{K}}$ for the spin-1/2 XY and the tricritical spin-1 BC quantum chains. For the XY chain, $S_{\mathcal{K}}$ saturates to $\ln{2}$ as predicted by U(1)$_4$ CFT, and agrees perfectly with the analytical solution [Fig. \ref{Fig:KdiffM} (c)]. Also in Fig. \ref{Fig:KdiffM}(b), we show the spin-1 BC chain results at the tricritical point, where the large-scale simulations using LTRG and SETTN mutually agree and are both in very good accordance with the CFT prediction $0.854258$. 
The ED spectra of the XY and spin-1 BC chains reveal that the primary fields are at $q=0$ and all relevant CFT fields are of the same parity, supporting the large-scale TTN results. [Sec. III of Supplementary Material].

\begin{figure}[tbp]
  \includegraphics[angle=0,width=1\linewidth]{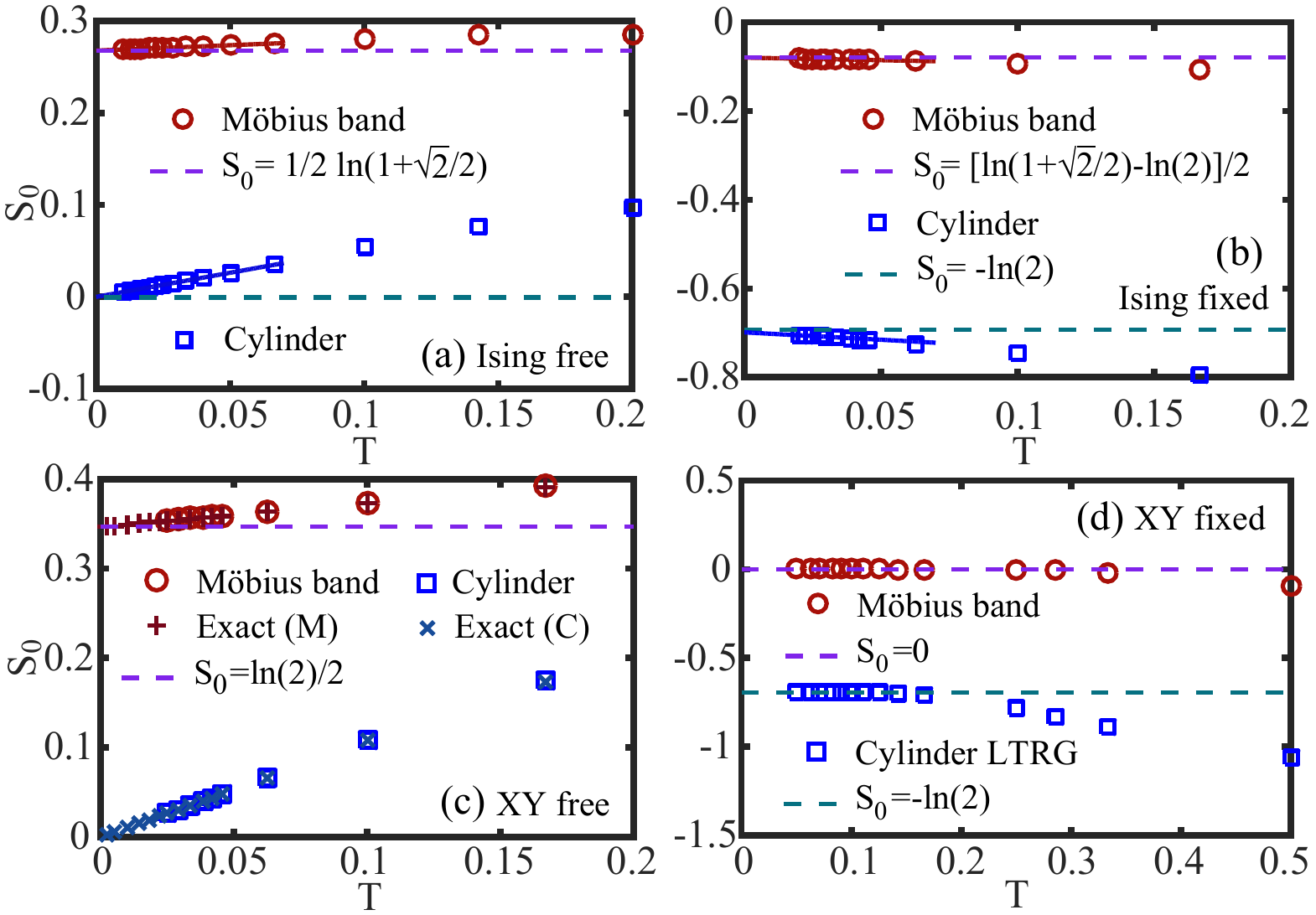}
  \caption{(Color online) Residual entropies of the TFI model are (a) $S_0=0$ for free and (b) $-0.693(3)\simeq-\ln{2}$ for fixed boundaries, based on the data of $L=350,400,450,500$. Correspondingly, on the M\"obius strip, $S_0$ extrapolates to 0.2675(2) and -0.0793(3). The lines in (a,b) are linear fittings of low-$T$ $S_0$ data. (c) The XY chain with free boundary, both numerical (fitted from $L=120,140,160$ data) and analytical results (with $L$ up to 3500) are shown, the cylindrical entropy vanishes while the best estimation on the M\"obius strip is $S_0=0.346572(1)\simeq \ln(2)/2$. (d) The XY chain with fixed boundary, $S_0=-0.6936(2)$ (cylinder) and $S_0 \simeq 4.5\times 10^{-4}$ (M\"obius), extrapolated from data with $L=160 \sim 220$.}
\label{Fig:OBC}
\end{figure}

\textit{Cylinder and M\"obius-strip worldsheets}.--- Next, we discuss the cylinder and the M\"obius partition functions. Residual entropies of the TFI and the XY OBC chains are shown in Fig. \ref{Fig:OBC}, with free and fixed boundaries, respectively. For AL entropy $S_{\rm{AL}}=\ln{g}$ in the TFI case,  $g=1$ for free boundaries, and $g=g_1 g_2=1/2$ for fixed boundaries on both ends. Due to the open boundaries, two linear fittings are performed to extract $S_0$: extrapolate $\ln{\mathcal{Z^{C,M}}}$ versus $L$ to evaluate the boundary partition functoin $\ln{z_b}(\beta)$ and then estimate the intercept by fitting $\ln{z_b}(\beta)$ by $-\beta e_s + S_0$. In Fig. \ref{Fig:OBC}(a), we show the extrapolated $S_0$ for the TFI on the cylinder and the M\"obius strip with free boundary. The cylinder intercepts vanish as $T \to 0$ (since $g=1$), while the M\"obius entropy converges to a nonzero universal value $S_{\mathcal{M}} = 0.2675(2) \approx \frac{\ln{(1+\sqrt{2}/2)}}{2}$, exactly one-half of $S_{\mathcal{K}}$ on the Klein bottle. For fixed boudary (i.e., switching on boundary magnetic field $h_b S_{1(L)}^x$, with $h_b=10$ in practice), we observe that $S_{\rm{AL}}=-\ln{2}$ [Fig. \ref{Fig:OBC}(b)], and the M\"obius entropy $S_{\mathcal{M}} = -0.0793(3) \simeq [-\ln{2} + \ln{(1+\frac{\sqrt{2}}{2})}]/2$, i.e., $\frac{1}{2} (\ln{g}+\ln{k})$. 

In Figs. \ref{Fig:OBC}(c,d), residual entropies in the OBC XY chain are also investigated, where the cylinder and M\"obius partition functions have analytical expressions (with free conformal boundary). The XY chain is described by a free boson field with compactification radius $R=(4\pi)^{-1/2}$, and thus the factor $g=\pi^{-1/2}(2R)^{-1}=1$ (free) and $g=\pi^{1/2} R=\frac{1}{2}$ (fixed, with  $h_b S_{1(L)}^x$ term applied) \cite{Affleck-1998,Zhou-2006}. The exact data and numerical results are shown in Fig. \ref{Fig:OBC} (c), which agree very well with each other. Notably, the former provides our best estimate $S_{\mathcal{M}} = 0.346572(1)$ for the XY-chain M\"obius entropy, which equals $\frac{1}{2} \ln{2}$ up to the sixth digit. Furthermore, in Fig.~\ref{Fig:OBC}(d), under fixed boundary $\frac{1}{2} S_{\rm{AL}}= \frac{1}{2} \ln{g}=-\frac{1}{2} \ln{2}$, which exactly cancels the twist entropy ($\frac{1}{2} \ln{k} = \frac{1}{2} \ln{2}$) and results in an overall $S_{\mathcal{M}}=\frac{1}{2}(\ln{2} - \ln{2})=0$. 

Therefore, based on the high-precision numerics and analytical solutions, we propose that the M\"obius partition function of a critical system is
\begin{equation}
\ln{\mathcal{Z^M}} \simeq L(-\beta \epsilon_0+\frac{\pi c}{24 v \beta}) + \ln{z_b^m},
\label{Eq:ParFuncMobius}
\end{equation}
where $ \ln{z_b^m} = - \beta e_s + \frac{1}{2} (\ln{g} + \ln{k})$.
Notably, this formula has a straight interpretation in terms of topology: The $\frac{1}{2}$ factor before $\ln{g}$ can be related to the fact that there exists only one open edge in the M\"obius strip (instead of two edges on the cylinder), and the $\frac{1}{2}$ factor before $\ln{k}$ reflects that a Klein bottle consists of two M\"obius strips. To be more concrete, we can virtually ``cut" the Klein bottle along a vertical line [see Fig. \ref{Fig:TNGeom}(a)], and rejoin the two parts [shown in Fig. \ref{Fig:TNGeom}(j), note the horizontal flip of the lower piece] in a way such that the Klein surface is ``massaged" into a flat cylinder with two crosscap boundaries, each contributes one-half Klein entropy \cite{Wang-2017}. Similarly, the M\"obius strip, after ``cutting" along a central line, does not break into two disjoint pieces but becomes a flat cylinder of size $\frac{1}{2} L \times 2\beta$, as shown in Fig.\ref{Fig:TNGeom}(k). Only one open edge (contributing $\frac{1}{2} \ln{g}$ boundary entropy) and one crosscap boundary ($\frac{1}{2}\ln{k}$ twist entropy). In addition, the factor $1/24$ in the bulk term in Eq. (\ref{Eq:ParFuncMobius}) also has a simple topology origin: on the $\frac{1}{2} L \times 2\beta$ ``cylinder", the bulk CFT correction should be $\frac{\pi c}{6 v (2\beta)} \frac{L}{2} = \frac{\pi c}{24 v \beta} L$.

\begin{figure}[tbp]
  \includegraphics[angle=0,width=1\linewidth]{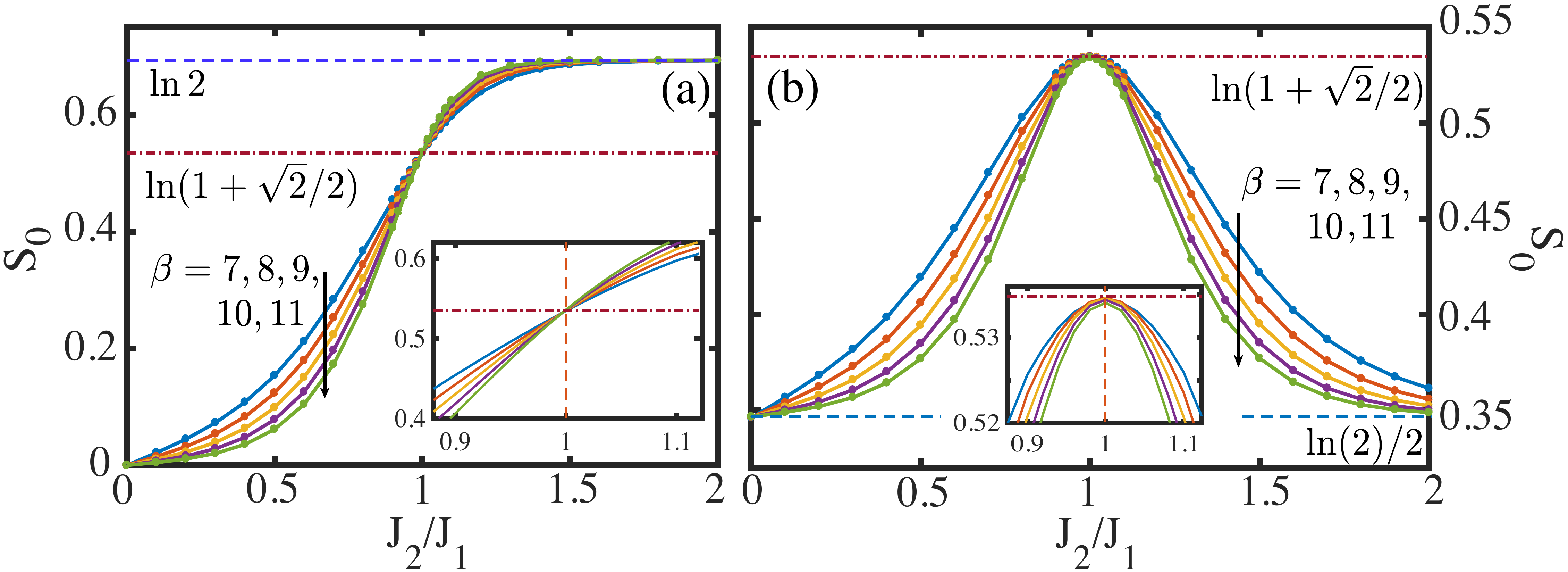}
  \caption{(Color online) Residual entropy $S_0$ versus $J_2/J_1$ in the single-chain Kitaev model for various temperatures, based on extrapolations of (a) $L=40, 60, 80$ data and (b) $L=42, 62, 82$ data. Insets zoom in the regime near the critical point $J_2/J_1=1$.}
\label{Fig:QPT}
\end{figure}

\textit{Accurate determination of quantum phase transition point}.--- One interesting application of the universal entropy is to pinpoint the QCPs from finite-temperature calculations, and it applies to both conventional symmetry-breaking quantum phase transitions and beyond.  In Fig. \ref{Fig:QPT} we demonstrate it with a Kitaev spin chain \cite{Liu-2012} $H_{\rm{KSC}} = -\sum_{n=1}^{L/2} J_1 S_{2n-1}^x S_{2n}^x +J_2  S_{2n}^y S_{2n+1}^y$, with a QCP at $J_2/J_1=1$, separating two disordered phases with distinct hidden string orders (while without any local order parameter) \cite{Zhang-2007}. In Fig. \ref{Fig:QPT} (a), we show the results of system sizes $L=4n$ (with integer $n$) and find $S_0$ changes from 0 (for $J_2<1$) continuously to $\ln{2}$ ($J_2>1$). $S_0$ converges rapidly to the universal value $\ln{(1+\frac{\sqrt{2}}{2})}$ at $J_2/J_1=1$, indicating an emergent Ising CFT, i.e., the same universality class as critical TFI. The $S_0-J_2$ curves in Fig.~\ref{Fig:QPT} show a cross point right at $J_2/J_1=1$ (the inset zooms in the regime near QCP).  For $L=4n+2$,  we observe that the peak of $S_0$ reaches the universal value $\ln{(1+\frac{\sqrt{2}}{2}})$ right at $J_2/J_1=1$ and singles out the QCP [see the inset of Fig.~\ref{Fig:QPT} (b)], while in two disordered phases $S_0 =  \ln{\sqrt{2}}$.  Therefore, we propose that the residual entropy remarkably serves as a tool to determine QCP from thermal date at relatively high temperatures, even when local order parameters are absent. Compared to standard techniques in quantum Monte Carlo (like computing Binder ratios), the universal entropy scheme achieves better accuracy, and demands significantly less resource.

Note the nonzero values of $S_0$ in both disordered phases are also very meaningful. They can be integer (2 for $J_2/J_1>1$ with $L=4n$) or even non-integer ($\sqrt{2}$ for both $J_2/J_1>1$ and $<1$ with $L=4n+2$), the latter reflects the ``topological" degeneracy on the Klein bottle related to Majorana zero modes \cite{FN:Majorana} living on the effective ``boundary" [Fig. \ref{Fig:TNGeom}(j)].

\textit{Conclusions and outlook}.--- We demonstrate, by thermal tensor network simulations, the existence of universal entropy corrections on nonorientable manifolds. The residual entropies can only select universal values related with the quantum dimensions of CFT primary fields , i.e., $S_{\mathcal{K}} = \ln{k}$ and $S_{\mathcal{M}}=\frac{1}{2}(\ln{k} + \ln{g})$ on the Klein bottle and the M\"obius strip, respectively. These novel universal entropies reflect non-integer ``groundstate" degeneracy of a critical quantum chain, and can be applied to accurately locate QCP from finite temperatures.

It is interesting to note that via AdS/CFT correspondence, the conformal TTN approaches, as well as the revealed universal entropies $S_{\mathcal{K}}$ and $S_{\mathcal{M}}$, might also find their applications in black hole entropy studies, where the holographic dual is right a thermal CFT \cite{Maldacena-2003,Shu-2016}. Last but not least, we notice that there are some recent works on partial reflection in symmetry-protected topological state to characterize topolgoical orders \cite{Pollmann-2012, Ryu-2017}, and expect that there also exist universal Klein or M\"obius (entanglement) entropies in the partially twisted ground states in quantum critical chains.

\begin{acknowledgments}
\textit{Acknowledgments}.--- The authors are indebted to Hong-Hao Tu for inspiring and intensive discussions, and also thank Jin Chen for useful discussions on CFT. This work was supported by the National Natural Science Foundation of China (Grant Nos. 11504014), the Research Fund for the Doctoral Program of Higher Education of China (Grant No. 20131102130005), and the Beijing Key Discipline Foundation of Condensed Matter Physics.
\end{acknowledgments}

\newpage
\mbox{}
\pagebreak
\widetext
\begin{center}
\textbf{\large Supplemental Materials: Conformal Thermal Tensor Networks and Universal Entropy on Topological Manifolds}
\end{center}
\setcounter{equation}{0}
\setcounter{figure}{0}
\setcounter{table}{0}
\setcounter{page}{1}
\makeatletter
\renewcommand{\theequation}{S\arabic{equation}}
\renewcommand{\thefigure}{S\arabic{figure}}
\renewcommand{\bibnumfmt}[1]{[S#1]}
\renewcommand{\citenumfont}[1]{S#1}

\section{Section I Finite-Temperature Renormalization Group Approaches}
\label{Sup:TTN}

In the numerical simulations of 1+1D quantum critical systems, we adopt two different thermal tensor network (TTN) techniques, namely, the linearized tensor renormalization group (LTRG) \cite{LTRG-SM,LTRG++-SM} and the series-expansion thermal tensor network (SETTN) \cite{SETTN-SM}. Here we briefly recapitulate these two approaches, as well as the their adaptation for the calculations on non-orientable manifolds like the Klein bottle or the M\"obius strip. In the following, we mainly focus on quantum chains with periodic boundary condition (PBC), i.e., the torus or the Klein bottle. The cylinder and M\"obius strip calculations are quite similar, and without special care taken on the long-range interaction term between the first and last sites due to PBC.

\begin{figure}[htb]
\includegraphics[angle=0,width=0.65\linewidth]{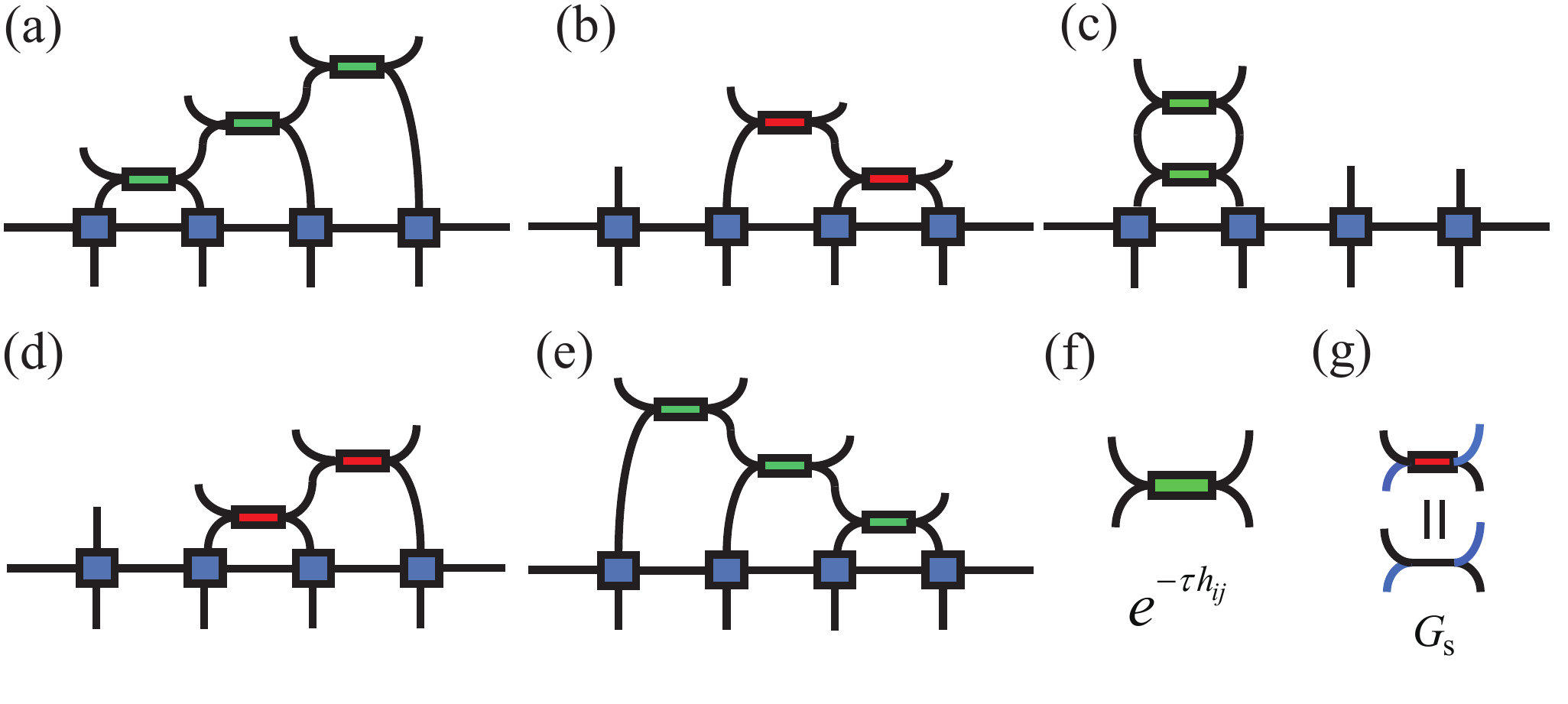}
\caption{LTRG approach for a finite-size quantum chain (of length $L$) with PBC, (a-e) accomplish a projection step of $e^{-2\tau H}$, where $H$ is the total Hamiltonian and $\tau$ is the Trotter slice. (a) Project local evolution gates $e^{-\tau h_{i,j}}$ [illustrated in (f)], site by site, to the density-matrix MPO to cool down the system. (b) Swap gate [plotted in (g)] is designed to exchange two upper physical indices of contiguous tensors, the successive applications of which will bring the physical index of the last site to the immediate right of the first one. (c) Project $e^{\tau h_{1,L}}$ twice to the first two tensors, and then (d) successively swap the physical index of the second site with the rest $L-2$ tensors to bring it back to the original position. (e) Perform the rest evolution operations and sweep from right to left to complete the $e^{-2\tau H}$ projection.}
\label{Fig:LTRG}
\end{figure}

The main procedure of LTRG is illustrated in Fig. \ref{Fig:LTRG} (see Ref. \onlinecite{LTRG++-SM} for more details on the finite-size LTRG and its bilayer generalization), where the themodynamics are calculated within a discrete Euclidean path integral via the Trotter-Suzuki decomposition \cite{Trotter-1959,Suzuki-1976}, and the density matrix (as a matrix product operator, MPO) can be evolved by projecting local evolution gates $e^{-\tau h_{ij}}$ [Fig. \ref{Fig:LTRG}(f)] successively to the MPO. Special treatment is needed for the PBC term, where one needs to swap a physical index from one end to the other [Fig. \ref{Fig:LTRG} (b)], perform ``local" projections [Fig. \ref{Fig:LTRG} (c)], and then swap it back [Fig. \ref{Fig:LTRG} (d)]. The swap gate $G_s$, as illustrated in Fig. \ref{Fig:LTRG} (g), exchanges two physical indices and is a unitary gate. Take the spin-1/2 chain as an example, $G_s$ is a four-by-four matrix with nonzero elements $\langle \uparrow_1 \uparrow_2 | G_s | \uparrow_1 \uparrow_2 \rangle = \langle \downarrow_1 \downarrow_2 | G_s | \downarrow_1 \downarrow_2 \rangle = \langle \uparrow_1 \downarrow_2 | G_s | \downarrow_1 \uparrow_2 \rangle =  \langle \downarrow_1 \uparrow_2 | G_s | \uparrow_1 \downarrow_2 \rangle = 1$. Through the imaginary-time evolution, we obtain the matrix product density matrix at inverse temperature $\beta$. To obtain the partition function on the torus or the Klein-bottle manifold, one can perform a direct [Fig. 1 (g)] or a twisted trace [Fig. 1 (i)].

\begin{figure}[htb]
\includegraphics[angle=0,width=0.4\linewidth]{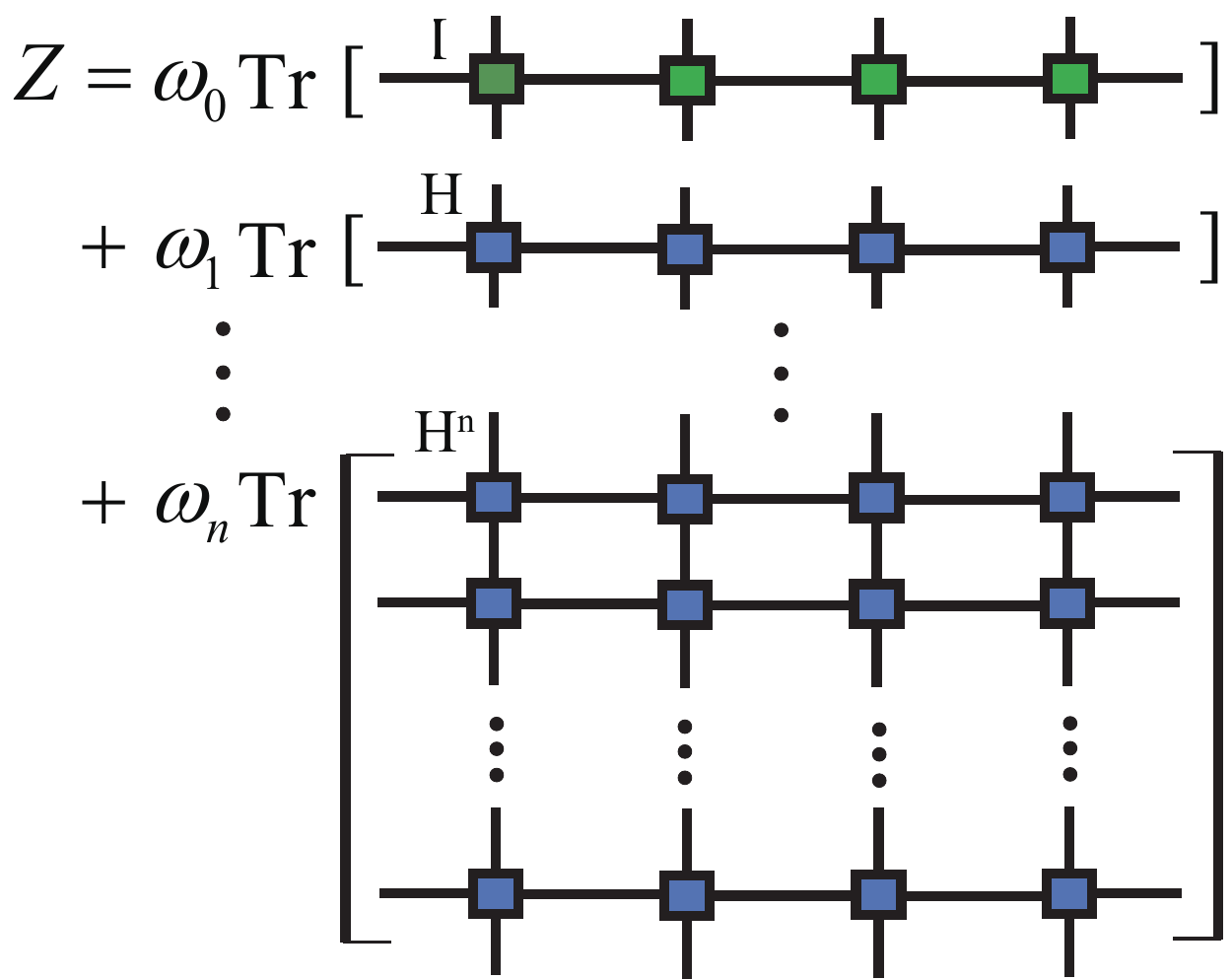}
\caption{Series-expansion thermal tensor network method, where the Hamiltonian and its power $H^n$ are expressed in terms of MPOs. A direct trace of the tensor network consisting of MPOs corresponds to the partition function on the torus or cylinder manifold, and a twisted trace leads to the Klein-bottle or M\"obius-strip partition function. The partition functions are obtained by a weighted sum of these tensor-network trace, with weights $\omega_n = (-\beta)^n/n!$.}
\label{Fig:SETTN}
\end{figure}

Besides the LTRG calculations, we also adopt the series-expansion thermal tensor network (SETTN), which employs the Taylor expansion of density matrix $e^{-\beta H}$ and is essentially free of Trotter error \cite{SETTN-SM}. The main idea is illustrated in Fig. \ref{Fig:SETTN}, and note that the calculation of partition function on orientable (non-orientable) manifolds amounts to a series of direct(twisted) tensor traces of $H^n$ terms: $\ln{Z}(\beta) = \sum_{n=0}^{N} \omega_n(\beta) \rm{Tr}(\mathcal{O} H^n)$, where $\mathcal{O} $ is identity (spatial reflection $\mathcal{P}$) for the direct(twisted) trace, and $\omega$ obeys the Poisson distribution.

SETTN is very accurate and flexible algorithm for finite-temperature calculations, and the long-range interaction term can also be treated very conveniently \cite{Frowis-2010-SM,Pirvu-2010-SM}. In the present case, there exists a long-range coupling due to PBC, which need to be encoded in the MPO representation of the total Hamiltonian. Take the transverse-field Ising (TFI) model $H = \sum_i (-S^z_i S^z_{i+1} + h S^x_i)$ as an example, we need the following tensor $A$ to express an OBC Hamiltonian as a translation-invariant MPO: $A_{1,1}^{s,d} = \mathbb{I}^{s,d}, A_{1,2}^{s,d} = (S^z)^{s,d}, A_{2,3}^{s,d} = (S^z)^{s,d}, A_{3,3}^{s,d} = \mathbb{I}^{s,d}$, and $A_{1,3}^{s,d} = h (S^x)^{s,d}$, where $s,d$ run over the local physical dimension. Since the quantum chain is with OBC, $A$ tensor at the first site has a single left index (chosen as ``1"), and the last tensor has a single right index (fixed as ``3"). For PBC Hamiltonian, an additional channel is needed: $A_{1,2'}^{s,d} = (S^z)^{s,d}$ on the first site, $A_{2',3}^{s,d} = (S^z)^{s,d}$ on the last site, and $A_{2',2'}^{s,d} = \mathbb{I}^{s,d}$ for the rest sites to switch on the PBC channel. Thus the MPO representation under PBC has one more bond dimension than that in the OBC case.

MPO representations for the other two models considered in the present work are as follows. For PBC XY chain, MPO representation is of bond dimension $\chi=6$, with $A$ tensors: $A_{1,1}^{s,d} = \mathbb{I}^{s,d}, A_{1,2}^{s,d} = (S^x)^{s,d}, A_{2,4}^{s,d} = -(S^x)^{s,d}, A_{1,3}^{s,d} = (S^y)^{s,d}, A_{3,4}^{s,d} = -(S^y)^{s,d}, A_{4,4}^{s,d} = \mathbb{I}^{s,d}$, and PBC channel $A_{1,2'}^{s,d} = (S^x)^{s,d}, A_{1,3'}^{s,d} = (S^y)^{s,d}$ on the first site, $A_{2',4}^{s,d} = -(S^x)^{s,d}, A_{3',4}^{s,d} =-(S^y)^{s,d}$ on the last site, and $A_{2',2'}^{s,d} = A_{3',3'}^{s,d} = \mathbb{I}^{s,d}$ on the rest sites. For the spin-1 BC chain (PBC), we have a $\chi=4$ MPO which is very much like that of the TFI chain, but with the difference that $ A_{1,3}^{s,d} = [-h S^x - D(S^z)^2]^{s,d}$, and $\{S^x,S^y,S^z\}$ are spin-1 operators now.

\begin{figure}[!htb]
\includegraphics[angle=0,width=0.75\linewidth]{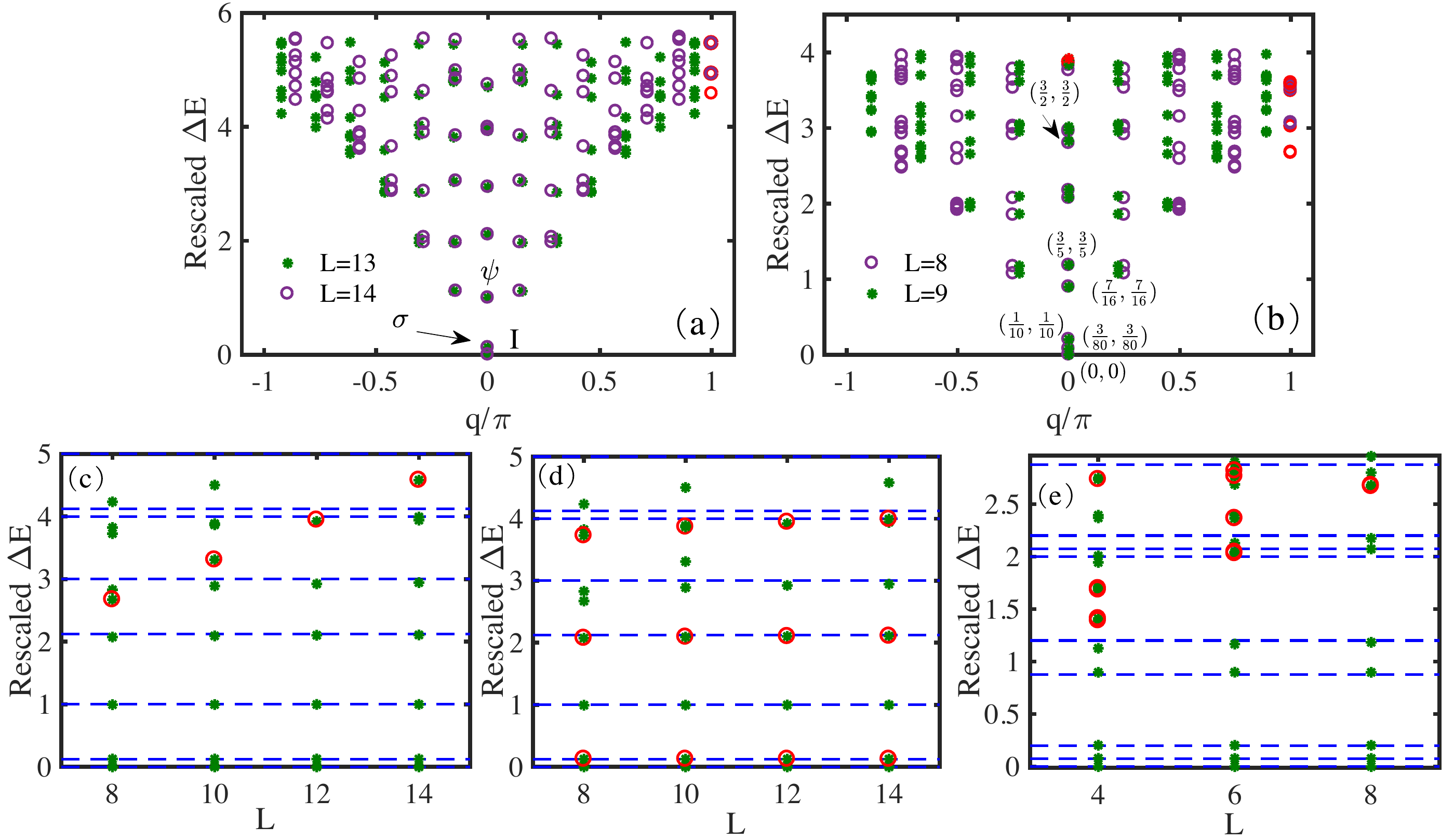}
\caption{(Color online) (a) ED spectra of the FM Ising chains of even and odd lengths, where similar Virasoro tower structures appear in both cases, i.e. all primary fields are of parity $+1$, as well as corresponding descendants, at momentum $q=0$. There exist some high-energy states (with constant energy) with odd parity ($P=-1$) located at momenta $q=\pi$ (for even length). (b) ED spectra of tricritical Ising point in the spin-1 Blume-Capel chain. (c) Rescaled excitation energy $\Delta E$ of FM Ising model, where rescaled energies of the CFT fields stay on a horizontal line, while the constant-energy states (highlighted by circles) remain in the high-energy sector and thus $\Delta E$ increases almost linearly with $L$. For AF Ising chain in (d), instead we observe the odd-parity states are CFT fields and thus contribute to low-$T$ partition function and affect the residual entropy. In (e) the tricritical spin-1 Blume-Capel model, we also observe the odd-parity states in the high energy sectors. }
\label{Fig:supED}
\end{figure}

\section{Sec. II Analytical solutions of the XY quantum chain on the cylinder and the M\"obius manifolds}
\label{Sup:Analytic}
In this section, we provide the analytical expressions of XY-chain partition functions on various manifolds including the cylinder and the M\"obius strip.

\subsection{A. cylinder partition function}
For OBC Hamitonian
\begin{equation}
\begin{aligned}
H_{\rm{XY}} & = - \sum_{j=1}^{L-1}(S_j^x S_{j+1}^x + S_j^y S_{j+1}^y) \\
& = - \frac{1}{2} \sum_{j=1}^{L-1}(S_j^+ S_{j+1}^- + S_{j+1}^+ S_j^- ) \\
& = - \frac{1}{2} \sum_{j=1}^{L-1}(c_j^{\dagger}c_{j+1} +c_{j+1}^{\dagger}c_{j}),
\end{aligned}
\end{equation}
where the Jordon-Wigner transformation
\begin{equation}
S_j^+ = c^\dagger_j e^{-i\pi \sum_{l=1}^{j-1} c_l^\dagger c_l}
\end{equation}
and
\begin{equation}
S_j^- = {c_j} e^{i \pi \sum_{l=1}^{j-1} c_l^\dagger c_l},
\end{equation}
is exploited.

We apply the stationary wave expansion
\begin{align}
c_j &= \sqrt{\frac{2}{L+1}} \sum_{k=1}^{L}d_k \sin(\frac{k \pi}{L+1}j),\\
d_k &= \sqrt{\frac{2}{L+1}} \sum_{j=1}^{L}c_j \sin(\frac{k \pi}{L+1}j),
\end{align}
to express the Hamiltonian as a free fermions chain as
\begin{equation}
H=-\sum_{k=1}^{L} \epsilon_k d_k^{\dagger}d_k,
\end{equation}
where the dispersion relation $\epsilon_k = -\cos(\frac{k\pi }{L+1})$. Straightforwardly, the cylinder partition function of the XY chain is
\begin{equation}
\begin{aligned}
Z^{\mathcal{C}} &=\mathrm{Tr}(e^{-\beta H})\\
&=\prod_{k=1}^{L}(1+e^{-\beta \epsilon_k})
\end{aligned}
\end{equation}

\subsection{B. M\"obius-strip partition function}
The M\"obius-strip partition function can be evaluated following similar line developed in the cylinder case, while we need to consider the effect of spatial reflection operator ($\mathcal{P}$) on fermion operators $d_k^\dagger$.
\begin{equation}
\begin{aligned}
\mathcal{P} d_k^\dagger \mathcal{P}^{-1} & = \sqrt{\frac{2}{L+1}} \sum_{j=1}^L \mathcal{P} c_j^\dagger \mathcal{P}^{-1} \sin(\frac{k \pi}{L+1})\\
&=\sqrt{\frac{2}{L+1}} \sum_{j=1}^{L}c_{L-j+1}^{\dagger}Q\sin(\frac{k\pi}{L+1}j)\\
&=\sqrt{\frac{2}{L+1}} \sum_{j'=1}^{L}c_{j'}^{\dagger}Q \sin[{\frac{k \pi}{L+1}(L-j'+1)}]\\
&=\sqrt{\frac{2}{L+1}} \sum_{j'=1}^{L}c_{j'}^{\dagger}Q (-1)^{k+1} \sin(\frac{k \pi}{L+1}j')\\
&=(-1)^{k+1} d_k^\dagger Q.
\end{aligned}
\end{equation}
Then we consider the effect of $\mathcal{P}$ on a many-fermion state in the occupation number representation
\begin{equation}
\begin{aligned}
\mathcal{P} |n_1,n_2,\cdots,n_L\rangle &= \mathcal{P}(d_{1}^\dagger)^{n_{1}} (d_{2}^\dagger)^{n_{2}} \cdots (d_{L}^\dagger)^{n_{L}} |\rm{vac}\rangle\\
&= \mathcal{P}(d_{1}^\dagger)^{n_{1}}\mathcal{P}^{-1}\mathcal{P} (d_{2}^\dagger)^{n_{2}}\mathcal{P}^{-1} \cdots \mathcal{P}(d_{L}^\dagger)^{n_{L}}\mathcal{P}^{-1}\mathcal{P} |\rm{vac} \rangle\\
&=[(-1)^{1+1}]^{n_1}(d_1^{\dagger})^{n_1}Q^{n_1} [(-1)^{2+1}]^{n_2}(d_2^{\dagger})^{n_2}Q^{n_2} \cdots [(-1)^{L+1}]^{n_L}(d_L^{\dagger})^{n_L}Q^{n_L}|\rm{vac} \rangle\\
&=[\prod_{k=1}^L(-1)^{n_k(k+1)}]\cdot (-1)^{N(N-1)/2} |n_1, n_2, \cdots, n_L \rangle,
\end{aligned}
\end{equation}
where the factor $(-1)^{N(N-1)/2}$ appears since we bring all parity operators $Q$ across $\{d^{\dagger}\}$ operators to right before the vaccumm ($Q=1$ there), and $n_k=0,1$ with $k=1,\cdots,L$. Therefore, the partition function can be evaluated as

\begin{equation}
\begin{aligned}
Z^{\mathcal{M}} &= \mathrm{Tr}(\mathcal{P} e^{-\beta \sum_{k=1}^{L}\epsilon_k d_k^\dagger d_k})\\
&=\sum_{\{n_k\}}\langle n_1,n_2,\cdots,n_L|\prod_{k=1}^{L} e^{-\beta \epsilon_k d_k^\dagger d_k} \mathcal{P} |n_1,n_2,\cdots,n_L \rangle \\
&=\sum_{\{n_k\}}[\prod_{k=1}^{L}e^{-\beta\epsilon_k n_k}(-1)^{n_k(k+1)}(-1)^{N(N-1)/2}],
\end{aligned}
\end{equation}
where $\sum_{\{n_k\}}$ means summing over all possible occupations of $n_k$. To simplify the above equation, we notice that if $N=\rm{odd}$, $(-1)^{N(N-1)/2}=(-1)^{(N-1)/2}$, while if $N$ is even $(-1)^{N(N-1)/2}=(-1)^{N/2}$.
\begin{equation}
\begin{aligned}
Z^{\mathcal{M}} = \sum_{\{n_k\}}^{N=\rm{even}}(\prod_{k=1}^Le^{-\beta \epsilon_k n_k}(-1)^{n_k(k+1)} (-1)^{N/2})+ \sum_{\{n_k\}}^{N=\rm{odd}} (\prod_{k=1}^Le^{-\beta \epsilon_k n_k}(-1)^{n_k(k+1)}(-1)^{(N-1)/2})
\end{aligned}
\end{equation}
We can calculate the partition functions in even and odd sectors, 
\begin{equation}
\begin{aligned}
Z^{\mathcal{M}}_{\rm{even}} &= \sum_{\{n_k\}}^{N=\rm{even}}(\prod_{k=1}^Le^{-\beta \epsilon_k n_k}(-1)^{n_k(k+1)} (-1)^{N/2})\\
&=\frac{1}{2}[\sum_{\{n_k\}}^{N=\rm{even}}\prod_{k=1}^Le^{-\beta \epsilon_k n_k}(-1)^{n_k(k+1)}(-1)^{N/2} + \sum_{\{n_k\}}^{N=\rm{odd}} \prod_{k=1}^Le^{-\beta \epsilon_k n_k}(-1)^{n_k(k+1)} (-1)^{N/2} \\
&\qquad + \sum_{\{n_k\}}^{N=\rm{even}} \prod_{k=1}^Le^{-\beta \epsilon_k n_k}(-1)^{n_k(k+1)}(-1)^{N/2} - \sum_{\{n_k\}}^{N=\rm{odd}} \prod_{k=1}^Le^{-\beta \epsilon_k n_k}(-1)^{n_k(k+1)} (-1)^{N/2}]\\
&=\frac{1}{2}[\sum_{\{n_k\}} \prod_{k=1}^Le^{-\beta \epsilon_k n_k}(-1)^{n_k(k+1)}(-1)^{n_k/2} + \sum_{\{n_k\}} \prod_{k=1}^Le^{-\beta \epsilon_k n_k}(-1)^{n_k(k+1)}(-1)^{n_k/2}(-1)^{n_k}]\\
&=\frac{1}{2}[\prod_{k=1}^{L}(1+i(-1)^{k+1}e^{-\beta \epsilon_k})+\prod_{k=1}^{L}(1-i(-1)^{k+1}e^{-\beta \epsilon_k})],
\end{aligned}
\end{equation}
where we have employ the trick that the sum of the even and odd $N$ sectors recover an uncontraint sum and thus we can exchange the order of $\sum_{\{n_k\}}^{N}$ and $ \prod_{k=1}^L$. Note that $(-1)^{1/2} = i$ and $(-1)^{-1/2}=-i$, similarly we get
\begin{equation}
\begin{aligned}
Z^{\mathcal{M}}_{\rm{odd}}&=\sum_{\{n_k\}}^{N=\rm{odd}} \prod_{k=1}^Le^{-\beta \epsilon_k n_k}(-1)^{n_k(k+1)}(-1)^{(N-1)/2}\\
&=-\frac{i}{2} [\prod_{k=1}^{L}(1+i(-1)^{k+1}e^{-\beta \epsilon_k})-\prod_{k=1}^{L}(1-i(-1)^{k+1}e^{-\beta \epsilon_k})].
\end{aligned}
\end{equation}
Finally, the M\"obius-strip partition function is
\begin{equation}
\begin{aligned}
Z^{\mathcal{M}} &= Z^{\mathcal{M}}_{\rm{even}} + Z^{\mathcal{M}}_{\rm{odd}} \\
&=\frac{1}{2}[\prod_{k=1}^{L}(1+i(-1)^{k+1}e^{-\beta \epsilon_k})+\prod_{k=1}^{L}(1-i(-1)^{k+1}e^{-\beta \epsilon_k})]+\\
&\quad-\frac{i}{2} [\prod_{k=1}^{L}(1+i(-1)^{k+1}e^{-\beta \epsilon_k})-\prod_{k=1}^{L}(1-i(-1)^{k+1}e^{-\beta \epsilon_k})]\\
&=\frac{1-i}{2}\prod_{k=1}^{L}(1+i(-1)^{k+1}e^{-\beta \epsilon_k})+\frac{1+i}{2}\prod_{k=1}^{L}(1-i(-1)^{k+1}e^{-\beta \epsilon_k})
\end{aligned}
\end{equation}

\section{Sec. III Energy spectra and eigenstate parity}
\label{Sup:Spectrum}
In this section, we provide more details on energy spectra obtained from exact diagonalization (ED). Note that the parity ($\mathcal{P}$) and the momentum ($\mathcal{T}$) operators both commute with the Hamiltonian $H$, however they do not commute with each other and the eigenstates might not have well-defined parity and momentum at the same time. We diagonalize $H$ in such a way that every eigenstate $|e_q \rangle$ has a well-defined energy $\epsilon$ and momentum $q$ [as shown in Fig. \ref{Fig:supED}(a)], and evaluate the parity by calculating the expectation value $P = \langle e_q | \mathcal{P} | e_q \rangle$ in the common eigenvectors of $\{ H, T \}$.

Fig. \ref{Fig:supED}(a,b) displays the spectra of the ferromangetic TFI model and the spin-1 Blume-Capel chain at the tricritical point, where the spectra of even- and odd-site chains show similar behaviors: $P=1$ for all primary and descendant fields at $q=0$ sector, and $P=0$ for quantum states with $0<q<\pi$. In the FM Ising case, the symmetric states entering the Klein-bottle partition function are of positive parity at $k=0$, and the negative-parity states at $k=\pi$ (marked by red) do not contribute due to its constant energy even in $L\to \infty$. To show these states as high-energy states and are irrelevant in low-temperature properties, we plot the excitation energy ($\Delta E = E-E_g$, $E_g$ is the groundstate energy) in Figs. \ref{Fig:supED}(c). In the plot, there exist two kinds of states showing distinct behaviors as $L$ enlarges, i.e., CFT fields with excitation energy decreasing inversely proportional with $L$ [$\Delta E=\frac{2\pi v}{L}(h+\bar{h}+n+\bar{n})$] and states with constant excitation energy (remaining in the high energy sectors as the circles indicate). The contributions of the latter to low-temperature properties, which we are only interested in, will be exponentially suppressed and are thus irrelevant as $T \to 0$. Figure \ref{Fig:supED}(d) shows the rescaled excitation energies of the AF Ising (of even length), where we see clearly that the odd-parity states are CFT fields. These states enter the Klein-bottle partition function, resulting in a different residual entropy, i.e., $S_0 = \ln{(1-\sqrt{2}/2)}$. In Fig. \ref{Fig:supED}(e), the rescaled  spin-1 Blume-Capel model at the tricritical point \cite{Blume-1966-SM,Capel-1966-SM} is shown, where the states with odd parity (red circles in Fig. \ref{Fig:supED}(b) are irrelevant high-energy states. 



\section{Sec. IV Accurate determination of quantum critical points by the Klein twist}

\begin{figure}[htb]
\includegraphics[angle=0,width=0.55\linewidth]{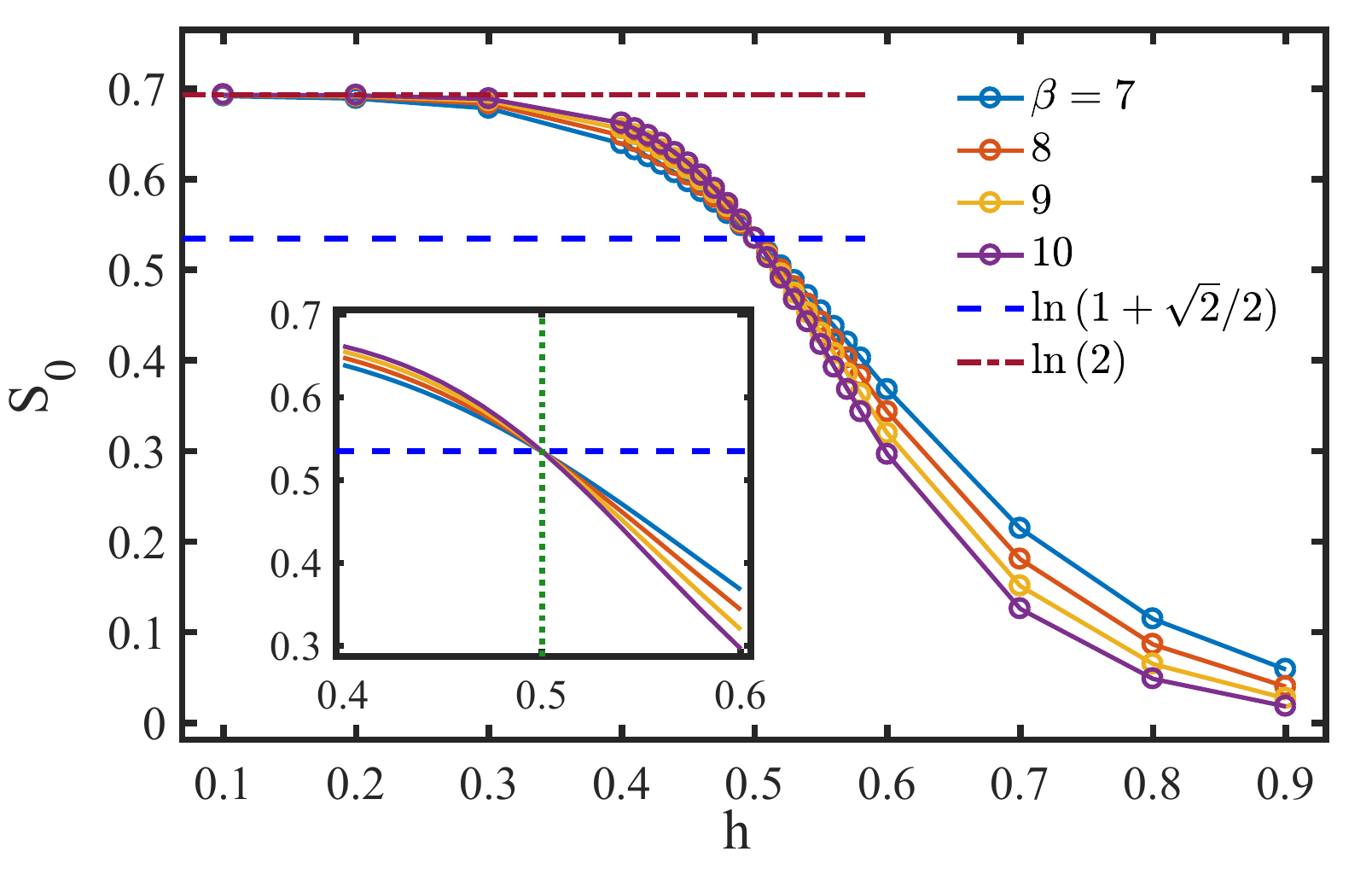}
\caption{(Color online) Residual entropy in the transverse-field Ising model with various fields $h$ and inverse temperatures $\beta$, the extrapolation of $S_0$ are based on small system sizes up to $L=80$. Inset zooms in $S_0$ in the regime $0.4 \leq h \leq 0.6$, where fast-converging cross points are observed. Lines in the inset are in the same color code as the main panel.}
\label{Fig:QCP}
\end{figure}

As a useful application, we use the Klein-bottle entropy to identify QCPs from thermal data. Firstly, we calculate the Klein-bottle entropy $S_{\mathcal{K}}$ of a transverse-field Ising model $H = \sum_i -S_i^x S_{i+1}^x - h S_i^z$ on the Klein-bottle worldsheet. Following the same computational scheme described in the main text, we linearly extrapolate $\ln{\mathcal{Z^K}}$ with system sizes $L$ and obtain the intercepts $S_0$ (shown in Fig. \ref{Fig:QCP}).

Note that in the calculations, we tune the transverse fields $h$ from 0.1 to 0.9, and put an emphasis in the regime ($0.4 \leq h \leq 0.6 $) near the known QCP $h_c=0.5$. In the ferromagnetic phase ($h \ll h_c=0.5$) $S_0$ approaches $\ln{2}$ which reflects the two-fold degeneracy in the magnetic ordered phase. As magnetic fields increase, and $S_0$ violates $\ln{2}$ due to finite value of $\beta$, and eventually converges to $\ln{1}=0$ deep in the paramagnetic phase (where the ground state is unique).

\begin{figure}[htb]
\includegraphics[angle=0,width=0.55\linewidth]{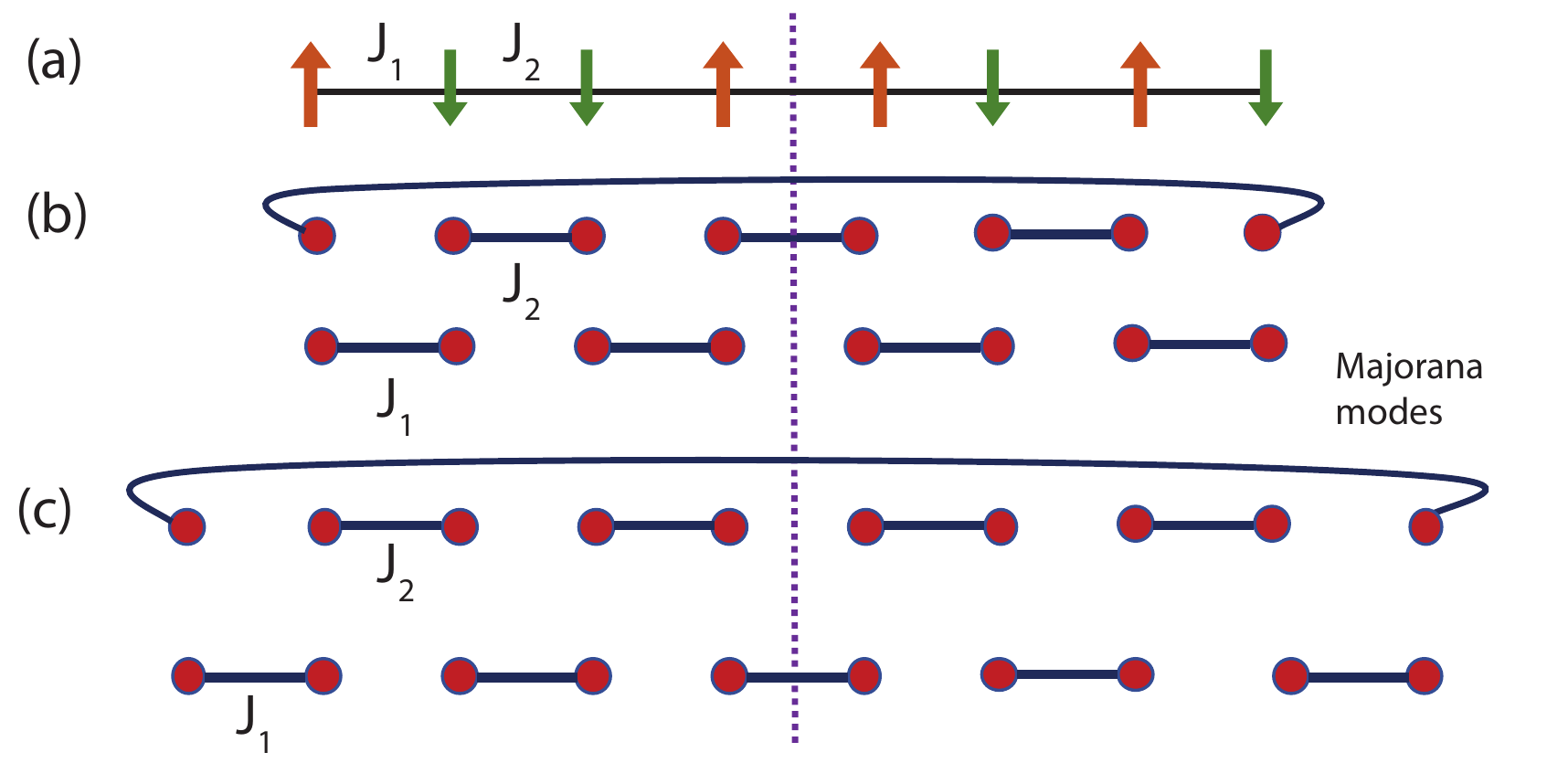}
\caption{(Color online) Degeneracy analysis of the (a) Kitaev spin chain in topological phases. (b) $L=4n$, the upper Majorana chain ($J_2$) contributes $S_0=\ln{2}$ while the $J_1$ chain contributes $S_0=0$; (b) $L=4n+2$, both Majorana chains produce $S_0=\ln{2}/2$ residual entropy.}
\label{Fig:TopoDegen}
\end{figure}

It is especially interesting to study the intercepts $S_0$ in cases with $h$ around the critical value $h_c$. Fig. \ref{Fig:QCP} shows that on the left-hand side ($h<h_c$) $S_0$ increases as temperatures get lower ($\beta$'s are larger), since the (two-fold) degenerate groundstate contributes greater to the residual entropy as the temperature decreases. According to similar arguments, we can understand that when $h>h_c$ the residual entropy $S_0$ gets smaller (and eventually $S_0=0$) as $T\to0$. At $h=h_c$, since $S_0$ converges quite fast versus $\beta$ and saturates to $S_{\mathcal{K}}=\ln{(1+\sqrt{2}/2)}$ as $\beta \geq 5$ [seen from Fig. 2 (c) in the main text]. Due to the different behaviors in the gapped and the critical regimes, we can pinpoint the QCP by the intersection point in $S_0$ curves. As shown in inset of Fig. \ref{Fig:QCP}, the intersection point converges very well even for even at intermediate temperatures $\beta = 7 \sim 10$. Intersection points provide a very good estimate of QCP, where $h_c$'s are estimated as $0.501, 0.5004, 0.50015$, from the pairwise $\beta=7,8$, $\beta=8,9$, and $\beta=9,10$ curves, respectively. Compared to Binder ratio method in quantum Monte Carlo, the system size $L$ and inverse temperature $\beta$ produce results with similar accuracy ($\sim$0.4985) needs to be as large as $\beta=L=80 \sim 100$, which demands clearly more computational resources. Therefore, we think this approach is a very practical way determining QCP without prior knowledge on the order parameter reflecting symmetry breaking, from relatively ``high"-temperature thermal data. 

Next, we consider the Kitaev spin chain model $H_{\rm{KSC}} = -\sum_{n=1}^{L/2} J_1 S_{2n-1}^x S_{2n}^x +J_2  S_{2n}^y S_{2n+1}^y$. As shown in Fig. 5 of the main text, we provide numerical results on the QPT between two disordered phases without any local order parameter, where the universal entropy can also be used to accurately pinpoint the QCP. The universal entropy $S_{\mathcal{K}}= \ln{(1+\sqrt{2}/2)}$ right at the QCP is the same as the TFI chain, reflecting the fact that this QPT belongs to the same universality class as the Ising transition. 

In addition, the residual entropies in the two disordered phases are also interesting, which reflects the ``topological" degeneracy related with Majorana zero modes. As shown in Fig.~\ref{Fig:TopoDegen} (a), the Kitaev spin chain can be mapped to a Majorana chain, coupled by $J_1$ ($J_2$) on odd (even) bonds. When the central line (about which the spatial reflection takes place) intersects a pairing bond, one gets additional $\ln{\sqrt{2}}$ residual entropy. In (b), when $J_1=0$ and $J_2=1$, the reflection line cuts two pairing bonds and results in a residual entropy of $\ln{2}$; while for $J_1=1$ and $J_2=0$, no paring bond is ``cut" and thus $S_0=0$. Therefore, this residual entropy counts the pairing bonds intersected by the reflection axis. This counting rule can be understood as following: although there exist no open edges which host true edge modes, the Klein bottle manifold has two effective boundaries due to the twist. When the reflection involves virtual edge modes  by ``cutting" the pairing bond, rejoining the worldsheet, and exposing the Majoranas to the effective boundaries [Fig. 1(j) in the main text], it contributes $\sqrt{2}$-fold degeneracy.  The degeneracy on the Klein bottle can be easily verified also in the spin representation in extreme cases (where only one coupling is nonzero): the total degeneracy is $\mathbb{D}=2^{(\frac{L}{2}-2)/2} \cdot 2^2 = 2^{L/4} \cdot 2$, leading to a residual entropy $\ln{2}$ when $J_1=0$ and $J_2 = 1$; while the total degeneracy is simply $\mathbb{D}=2^{(\frac{L}{2})/2}= 2^{L/4}$, with zero residual entropy for  $J_1 = 1$ and $J_2=0$.

Similarly, one can use this counting rule to correctly estimate the residual entropy in chains of size $L=4n+2$ in Fig.~\ref{Fig:TopoDegen}(c): When $J_1=0, J_2=1$ (or $J_1=1$, $J_2=0$), the residual entropy is $\frac{1}{2} \ln{2}$ due to the intersection of a single paring bond. In the spin representation, the the total degeneracy can be easily estimated as $\mathbb{D} = 2^{(\frac{L}{2}-1)/2} \cdot 2 = 2^{L/4} \cdot 2^{1/2}$, which also suggests a non-integer residual degeneracy.


\begin{thebibliography}{99}
\bibitem{FN:g1g2} $g=g_1 g_2$ reflects the non-integer groundstate degeneracy related with the left ($g_1$) and right ($g_2$) conformal boundaries, the same boundary condition is imposed on both ends and thus $g_1=g_2$ henceforth in this work.
\bibitem{FN-AFIsing} The contributions of symmetric states at $\pi$ (descendent of $\sigma$ fields) in the Klein bottle partition function is multiplied by a minus sign due to its negative parity.
\end{thebibliography}
\end{document}